\documentclass[12pt]{iopart}
\usepackage{amssymb}
\usepackage{epsfig}
\eqnobysec

\begin{document}

\title[Network Sensitivity]{Network Sensitivity to Geographical Configuration}

\author{Antony C Searle, Susan M Scott and David E McClelland}

\address{Department of Physics and Theoretical Physics, Faculty of Science,
The Australian National University, Canberra ACT 0200, AUSTRALIA}

\eads{\mailto{Antony.Searle@anu.edu.au}, \mailto{Susan.Scott@anu.edu.au},
\mailto{David.McClelland@anu.edu.au}}

\begin{abstract}
Gravitational wave astronomy will require the coordinated analysis
of data from the global network of gravitational wave
observatories. Questions of how to optimally configure the global
network arise in this context. We have elsewhere proposed a
formalism which is employed here to compare different
configurations of the network, using both the coincident network
analysis method and the coherent network analysis method.  We have
constructed a network model to compute a figure-of-merit based on
the detection rate for a population of standard-candle binary
inspirals. We find that this measure of network quality is very
sensitive to the geographic location of component detectors under
a coincident network analysis, but comparatively insensitive under
a coherent network analysis.
\end{abstract}

\submitto{\CQG} 

\maketitle

\section{Introduction}

Considerations of network analysis have helped shape the growth of
the global network of gravitational wave detectors. The sites of
the Laser Interferometric Gravitational-wave Observatory (LIGO)
detectors were chosen to facilitate a coincident network analysis
[1]; far enough apart to produce a detectable direction-dependent
relative time delay, but close enough together to have similar
antenna patterns. Likewise, the proposed site for the Australian
International Gravitational Observatory (AIGO) was chosen to be
near-antipodal to the LIGO detectors and thus to share their
antenna patterns [2]. Instruments such as VIRGO [3] and the
proposed Laser Cryogenic Gravitational Telescope (LCGT) [4] have
not been sited with regard to such considerations.

The coincident network analysis technique [5, 6], in its simplest
form, allows independent searches to be performed by each detector
in the network; a signal is only detected by the network when the
signal is detected by each member detector. A more recently
proposed technique is \emph{coherent} network analysis [7, 8],
whereby the output of all detectors is collected and then a single
search is performed on the combined data. The coherent network
analysis has a theoretical advantage over the coincident network
analysis, but the practical merits of each are still under debate.

The increasing viability of the new coherent network analysis
technique [7, 8] encourages us to reconsider existing results
about the global network; in particular, the influence of
instrument siting on the quality of the network as a whole [6].

We have introduced a formalism [9] to answer some of the questions
arising about how best to configure a network of gravitational
wave detectors; in particular, how to optimally site a new
detector augmenting an existing network, so as to maximize the
detection rate for a standard-candle binary inspiral, and whether
the effect is significant enough to warrant consideration in
planning the growth of the global network.

\section{Network Model}

Consider the parameter space representing a network of $n$
interferometric gravitational wave detectors.  The physical
location and orientation of an interferometer can be defined by
the latitude $\theta$ and longitude $\phi$ of the beam-splitter,
and the orientation angle $\psi$ of the x-arm with respect to
north, under the assumption of a horizontal detector on a
perfectly spherical Earth.  If the detectors are further assumed
to be identical up to location and orientation (that is, with
identical arm lengths and noise curves), then a network may be
described by the $3n$ parameters $[(\theta_1, \phi_1, \psi_1),
\ldots, (\theta_n, \phi_n, \psi_n)]$, forming a $3n$ dimensional
parameter space $\Theta$.

We have devised a \emph{figure of merit}, $f:\Theta \rightarrow
\mathbb{R}$, as a measure of the quality of the network.  We use a
quantity proportional to the detection rate for a population of
randomly oriented standard-candle binary inspirals uniformly
distributed in a flat space.  For a network with identical
detectors, uncorrelated noise and a fixed source class, the
structure of the noise curve and strain waveform contribute only a
constant multiplicative factor to the detection rate, so that we
may neglect them and consider only the relative orientation and
distance between the network and the source, and the analysis
algorithm used.  We follow Finn's maximum likelihood test
formulations of the coincident and coherent algorithms [7], and
additionally assume, for computational amenability, that the
threshold $\lambda$ is \emph{high} with respect to the noise power
(see [9] for details).

The $3n$-dimensional parameter space is too large to be computationally
amenable, and so we consider only proper subsets of particular interest. For
example, the best site (under our figure of merit) to augment an existing
network of $n$ detectors can be found by fixing the first $n$ detectors of a
$n+1$ detector network, and varying only $(\theta_{n+1}, \phi_{n+1},
\psi_{n+1})$.  Furthermore, noting that the figure of merit depends only weakly
on the orientation\footnote{See Figures \ref{lm}, \ref{ls}, \ref{hlvm},
\ref{hlvs}: at the upper and lower edges, corresponding to the north and
south geographic poles, the detectors rotate in place as the longitude
varies; despite this the figure of merit remains constant to a good
approximation.}, $\psi_{n+1}$, we can fix it to an arbitrary value, and vary
only the latitude and longitude. The figure of merit over this two-dimensional
section of parameter space then corresponds to a map of the relative merit of
different sites on the Earth for augmenting an existing network.

\section{Results}

\begin{figure}
\begin{center}
\epsfxsize=6in \epsfbox{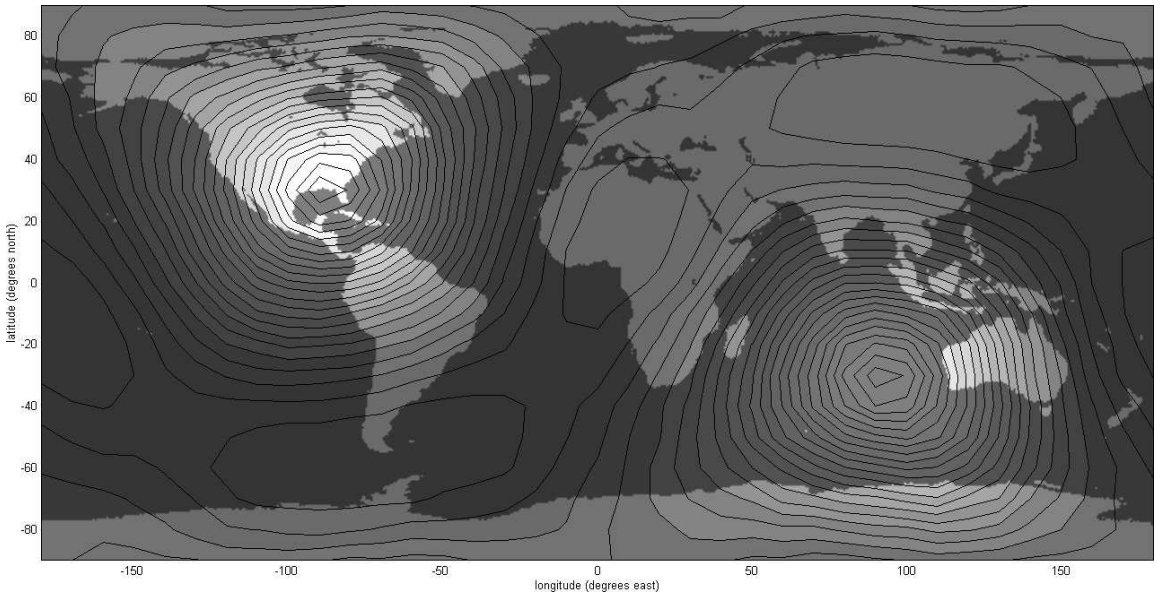}
\end{center}
\caption{Relative merit of an additional site to augment the LIGO Livingston
Observatory in a coincident analysis (lighter is better, contours every
2.5\%).
The minimum detection rate is 41\% of the maximum.}\label{lm}
\end{figure}

Consider first a single interferometer, at the site [10] of LIGO
Livingston Observatory (LLO). For a coincident network analysis,
the merit of an additional site to augment LLO is given in Figure
\ref{lm}.  It demonstrates, as expected, that sites near or
near-antipodal to LLO are best to augment it.  This is the
rationale behind the siting of the LIGO detectors, and the
proposed AIGO detector. The worst configurations produce a
substantially reduced detection rate; approximately 40\% that of
the optimal configuration. Unfortunately, the locations of VIRGO
and the proposed LCGT fall into this category.

It is interesting to note that for this simple case the map bears
some resemblance to the ``peanut'' antenna pattern of the fixed
single detector; the weak directionality of the varying detector,
and the superiority of a co-aligned network [11] are responsible
for this effect [9]. This resemblance breaks down for more
complicated networks.

\begin{figure}
\begin{center}
\epsfxsize=6in \epsfbox{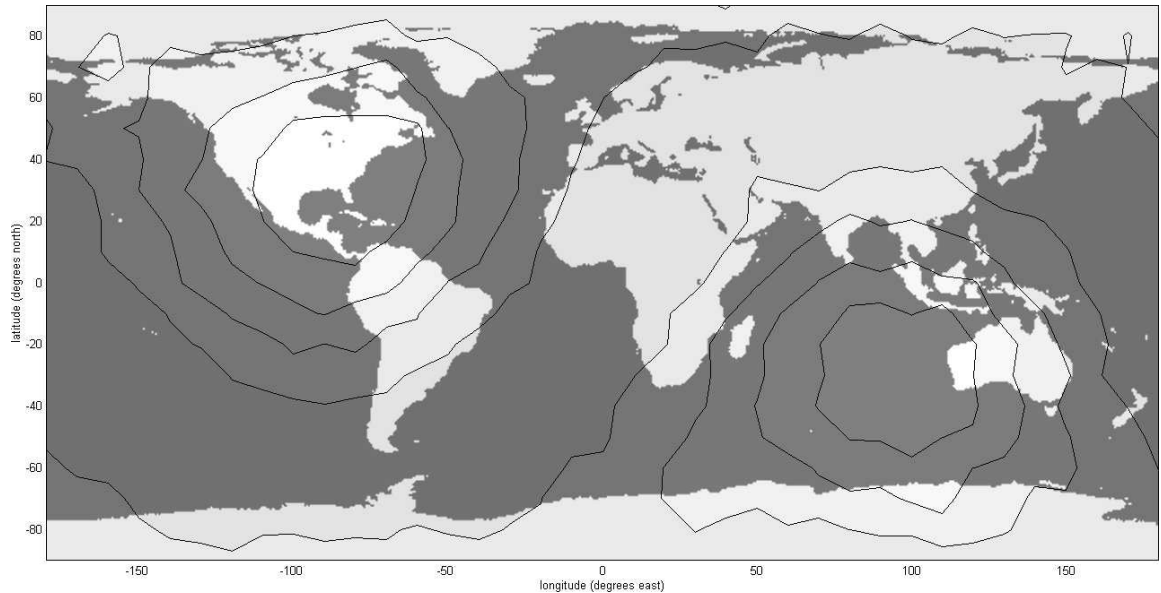}
\end{center}
\caption{Relative merit of an additional site to augment the LIGO Livingston
Observatory in a coherent analysis (lighter is better, contours every 2.5\%).
The minimum detection rate is 89\% of the maximum.}\label{ls}
\end{figure}

Considering the same configuration of a fixed LLO detector and a varying
detector with a coherent network analysis in Figure \ref{ls}, the qualitative
structure of the map is similar, but quantitatively it is quite different.
For
a coherent analysis, the worst configurations produce a detection rate that is
still 90\% of optimal; site merit does not vary substantially with location.

We now move on to consider an approximation to the existing global
network of the larger interferometric gravitational wave
detectors. We model the LIGO-VIRGO network as three identical
interferometers at the sites of LIGO Hanford Observatory (LHO),
LIGO Livingston Observatory and VIRGO [10].  Note that this model
neglects the 2 kilometre LHO instrument, and the differences
between the LIGO and VIRGO instruments.  Similarly, we augment
this three-detector network with a fourth (identical) detector at
different locations and compare the relative detection rates of
the resulting network.

\begin{figure}
\begin{center}
\epsfxsize=6in \epsfbox{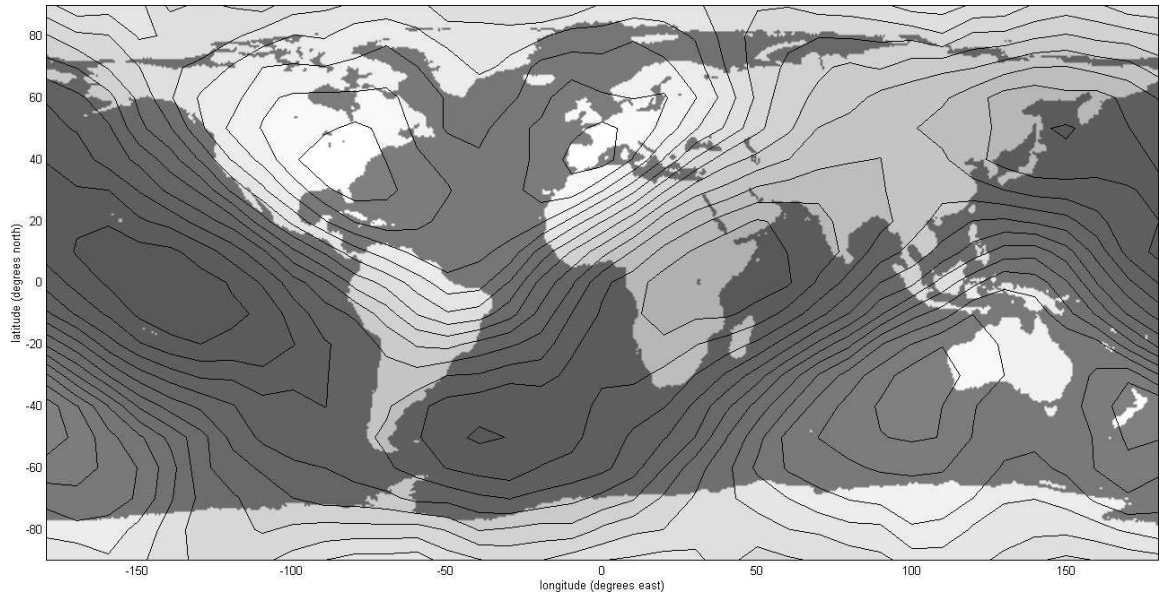}
\end{center}
\caption{Relative merit of an additional site to augment a network
consisting of
the LIGO Hanford (4km) Observatory, the LIGO Livingston Observatory and a 4km
LIGO I instrument at the VIRGO site, in a coincident analysis (lighter is
better, contours every 2.5\%). The minimum detection rate is 69\% of the
maximum.}\label{hlvm}
\end{figure}

Using a coincident network analysis in Figure \ref{hlvm}, we see that the
merit
of the network varies moderately with location, with multiple minima of about
70\% of the best achievable detection rates.

\begin{figure}
\begin{center}
\epsfxsize=6in \epsfbox{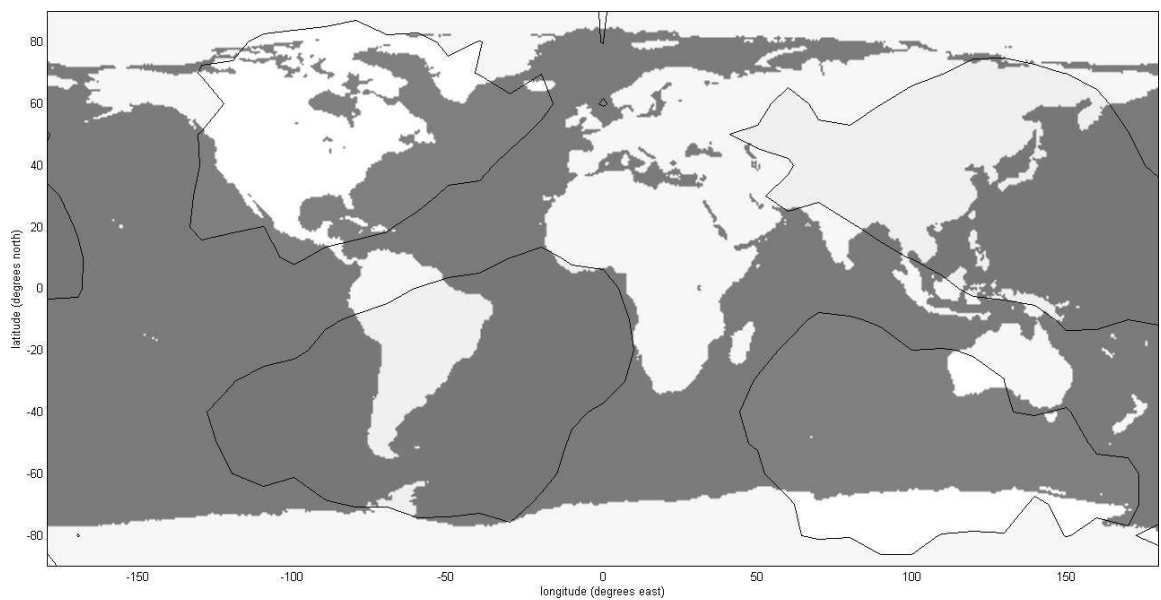}
\end{center}
\caption{Relative merit of an additional site to augment a network
consisting of
the LIGO Hanford (4km) Observatory, the LIGO Livingston Observatory and a 4km
LIGO I instrument at the VIRGO site, in a coherent analysis (lighter is
better, contours every 2.5\%). The minimum detection rate is 94\% of the
maximum.}\label{hlvs}
\end{figure}

Under a coherent network analysis in Figure \ref{hlvs}, we once again see a
qualitative similarity to Figure \ref{hlvm} in the locations of maxima and
minima, but quantitatively much less variation than in the coincident case,
with only 6\% separating the best and worst sites.  As expected, this
indicates
that AIGO is an optimal site to augment the existing global network; however,
the weak dependance of event rate on geographical location means that its
advantage over other sites is slight.

\section{Conclusion}

Under our model, it is clear that the (binary inspiral) detection rate for a
global network is insensitive to the geographical configuration of its
component detectors when a coherent analysis is used, in contrast to when a
simple coincident analysis is used. Whilst the LIGO detectors and proposed AIGO
detector are well sited to complement one another under a coincident
analysis, the sites of the VIRGO detector and proposed LCGT detector are
far from optimal; our results demonstrate that under a coherent analysis
the cost
of this sub-optimal siting is substantially reduced, on at least one figure of
merit.  In this sense, the global network is closer to optimal for a
coherent analysis than for a coincident analysis.  Our results also indicate
that since, under a coherent analysis, detection rate is insensitive to
detector siting, the location of an augmenting detector could be optimized for
other network properties (for example, directional resolution) without
compromising the event rate.

It is important to note that the model does not compare the absolute
detection rates for the two analysis techniques; we cannot say that one
method would produce a higher detection rate than the other for a given
false alarm rate.  We have considered only one class of source and one
figure of merit, but our formalism permits us to refine and extend the
network model and the range of questions with which it is used to answer.

\ack 

The authors wish to thank Lee Samuel Finn for many fruitful discussions on
this topic, and for making available computer systems supported by National
Science Foundation grant NSF 99-96213. This project is supported by an
Australian National University Research Grant, and by an award under the
Merit Allocation Scheme on the National Facility of the Australian
Partnership for Advanced Computing.

\end{document}